\documentclass[aps,prl,amsfonts]{revtex4}

\usepackage{graphicx}
\usepackage{graphics}
\usepackage{color}
\usepackage{bm}% bold math
\usepackage{multirow}
\usepackage{siunitx}
\usepackage{epsfig}
\usepackage{amsmath}

\newcommand{\E}{{\rm e}}

\newcommand{\I}{{\rm i}}

\newcommand{\D}{{\rm d}}

\newcommand{\beq}[1]{
	\begin{equation}
		\label{e#1} }
	
	\newcommand{\eeq}{
	\end{equation}
}
\begin{document}
	
\title{Simulation of ion demixing in halide perovskites using Cahn-Hilliard equation}

\author{V. Hol\'{y}$^{1,2}$, M. Dopita$^1$, L. Hor\'{a}k$^1$, J. Holovsk\'{y}$^3$}
\affiliation{$^1$Department of Condensed Matter Physics, Charles University, Ke Karlovu 5, 121 16 Prague 2, Czech Republic,}
\affiliation{$^2$Institute of Condensed Matter Physics, Masaryk University, Kotl\'{a}\v{r}sk\'{a} 2, 611 37 Brno, Czech Republic,}
\affiliation{$^3$Department of Electrotechnology, Technick\'{a} 2 Czech Technical University, 166 27 Prague 6, Czech Republic.}

%\author[1]{...}
%\author[2]{...}
%\author[1,3]{V. Hol\'{y}}
%\affil[1]{Department of Condensed Matter Physics, Charles University, Ke Karlovu 5, 121 16 Prague 2, Czech Republic} 
%\affil[2]{Department of Electrotechnology, Technick\'{a} 2 Czech Technical University, 166 27 Prague 6, Czech Republic}
%\affil[3]{Institute of Condensed Matter Physics, Masaryk University, Kotl\'{a}\v{r}sk\'{a} 2, 611 37 Brno, Czech Republic.}

%\date{\today}

\begin{abstract}
	Light-induced ion demixing in mixed-halide perovskites is simulated numerically using the phenomenological Cahn-Hilliard equation. In the model we consider the energy of local elastic deformation as well as the contribution of free carriers, assuming both the polaron and the bandgap-fluctuation models. The simulation shows that elastic deformation suppresses the demixing while free carriers promote it, however both effects lead to different ion distributions. The free-carrier-induced demixing appears only for larger starting random fluctuations of the ion concentration.	
\end{abstract}

\maketitle

\section{Introduction}

Mixed-halide perovskites (MHP) represent a new class of materials for photovoltaics. Their advantage lies in their inexpensive manufacturing \cite{SnaithJPCL2013} and very favorable optoelectronic properties, long free charge carrier diffusion lengths \cite{StranksSCI2013}, high absorption coefficient, sharp absorption edge onset \cite{deWolfJPCL2104}, and so-called defect tolerance \cite{SteirerACSEL2016,KimAEM2020}. Moreover, MHP materials exhibit a large bandgap tunability throughout the entire visible spectrum by varying the ratio between
different halide ions in the composition \cite{KitazawaJMS2002,FeldmannAOM2021},which makes these materials particularly promising for lighting \cite{TanNAT2014} or optimized tandem solar cells with silicon \cite{AlAashouriSCI2020}.

On the other hand, MHPs suffer from a partial demixing (phase segregation) during light soaking called the Hoke effect \cite{HokeCHS2015}. This operational instability has an unpleasant influence on the previously tailored absorption spectrum and further optoelectronic properties. Despite intense research, there is still no clear consensus on the origin of the halide ions migration and the forces acting against a concentration gradient. At least, various defects concentrations and some microstructure parameters correlated with segregation rates suggested several successful recipes for mitigation of ion demixing. However, to obtain reliable long-term operating devices, it is crucial to fully eliminate this effect. Therefore, it is necessary to identify all relevant migration channels.

Several explanations have been formulated for light-induced halide segregation. Polaron-induced strain gradients under illumination have been suggested to drive the nucleation of low-band gap iodine-rich domains \cite{BischakNANO2017}. Another model is based on band gap differences between perovskites with different halide compositions \cite{DragutaNatComm2017,SuchanAFM2022}; the band gap difference between mixed I/Br and I-rich domains, where photocarriers can reduce their free energy by funneling to the I-rich domains, is the driving force behind the demixing.

In this paper we develop a numerical mesoscopic model describing the kinetics of the segregation. The model is based on the phenomenological Cahn-Hilliard formalism \cite{CahnJCP1958,CahnAM1961,LeeCMS2014,KimMPE2016}. The values of the numerical parameters occurring in the formulas are taken from various literature resources and they are mostly obtained by atomistic ab-initio simulations. We use this model to investigate the kinetics of Br/I demixing in FA$_{1-y}$Cs$_y$PbBr$_x$I$_{1-x}$ perovskite compounds, where FA stands for a formamidinium cation CH(NH$_2$)$_2^+$. Typically $y=0.27$ and the atomic Br content $x$ (denoted by $c_\mathrm{Br}(\bm{r},t)$ in the following text) fluctuates in space and time.

\section{Cahn-Hilliard equation}

In this section we describe the light-induced decomposition of a FACsPbBrI MHP single crystal. The model is based on a numerical solution of the Cahn-Hilliard equation (CH):
\beq{1}
\frac{\partial\varphi}{\partial t}=M\Delta \left[\frac{\delta F(\varphi)}{\delta \varphi} -\kappa \Delta \varphi\right].
\eeq
Here $\varphi(\bm{r},t)=c_\mathrm{Br}(\bm{r},t)-c_0,\ c_0=\langle c_\mathrm{Br} \rangle$ is the order parameter, the time evolution of which we calculate, $c_\mathrm{Br}(\bm{r},t)$ is the local concentration of Br anions (normalized to unity), $c_0$ is its mean value, $M$ is the Br/I ion mobility, assumed not dependent on $\varphi$, $F(\varphi)$ is the Helmholtz free energy per one Br/I ionic site, $\delta F/\delta \varphi$ denotes the functional derivative, and $\kappa$ is a positive constant related to the energy penalizing steep gradients of $\varphi$ (``interface constant'').

The key issue of the model is the expression of the free energy $F$ and its derivative. In our model, the free energy consists of three terms
\beq{2}
F=F_\mathrm{mix}+E_\mathrm{deformation}+E_\mathrm{carriers},
\eeq
namely the free energy of mixing, the elastic energy density and the contribution of charge carriers generated by light absorption.

The free energy of mixing is calculated directly using a standard approach of thermodynamics of binary mixtures, the free energy per one Br/I ionic site reads \cite{Gaskell2003}:
\beq{3}
F_\mathrm{mix}(c_\mathrm{Br},T)=F_\mathrm{mix}^{(0)}+\Omega c_\mathrm{Br}(1-c_\mathrm{Br}) +k_B T \left[ c_\mathrm{Br} \ln c_\mathrm{Br} +(1-c_\mathrm{Br}) \ln(1-c_\mathrm{Br})\right],
\eeq
where the constant term $F_\mathrm{mix}^{(0)}$ disappears due to the derivation, $T$ is the absolute temperature and $\Omega = z[E_\mathrm{I,Br}-(E_\mathrm{I,I}+E_\mathrm{Br,Br})/2]$ is the solution constant ($z$ is the number of nearest Br/I neighbors, $E_\mathrm{A,B}$ is the interaction energy between neighboring ions of type A and B). Expressing the derivative of $F_\mathrm{mix}$ and using the order parameter, we obtain
\beq{4}
\frac{\delta F_\mathrm{mix}}{\delta \varphi}=\Omega(1-2c_0-2\varphi)+\frac{\partial \Omega}{\partial \varphi}(c_0+\varphi)(1-c_0-\varphi)
+k_B T \ln \frac{c_0+\varphi}{1-c_0-\varphi},
\eeq
where we have taken into account a possible dependence of the solution constant $\Omega$ on $\varphi$ (see below). It is worth mentioning that the series expansion of $\delta F_\mathrm{mix}/\delta \varphi$ for small $\varphi$ contains only odd powers of $\varphi$ if we neglect the slight $\Omega(\varphi)$ dependence and put $c_0=0.5$:
\beq{5}
\frac{\delta F_\mathrm{mix}}{\delta \varphi}=(4 k_BT -2\Omega)\varphi +\frac{16}{3}k_B T\varphi^3 +O(\varphi^5).
\eeq
If we ignore the elastic deformation and free carriers and use this power expansion, we obtain the Cahn-Hilliard equation in the standard form \cite{CahnJCP1958}
\beq{6}
\frac{\partial\varphi}{\partial t}=M\Delta \left(A_1 \varphi +A_3 \varphi^3 -\kappa \Delta \varphi\right),
\eeq
where $A_{1,3}$ are constants. The solution of the equation substantially depends on $A_1$. For $A_1<0$, i.e., for $\Omega >2k_BT$ the CH equation describes decomposition of the Br/I mixture, while for  $\Omega <2k_BT$ the Br/I fluctuations gradually disappear. This simple picture is substantially modified by elastic deformation and free carriers; we show in the next chapters that the elastic deformation inhibits the decomposition, while free carriers facilitate it.

The elastic-energy contribution to the free energy is calculated assuming linear elasticity and the validity of the linear Vegard's law $\epsilon_0=[a(c_\mathrm{Br})-a(c_0)]/a(c_0)=\eta\varphi$, see the details in Refs. \cite{ChenAMM1991,ZhuMSMSE2001,KimMPE2016}. Random fluctuations of the Br/I content give rise to a random deformation field, which is calculated assuming linear elasticity and neglecting the stress relaxation at the sample surface. The displacement field is obtained from the elastic equilibrium conditions
\beq{7}
C_{jkmn}\frac{\partial^2 u_m(\bm{r})}{\partial x_k \partial x_n}+f_j(\bm{r})=0,\ j,k,m,n,=1,2,3;\ \bm{r}=(x_1,x_2,x_3),
\eeq
where $C_{ijkl}$ are the elastic constants of MHP in the 4-index notation (assumed not dependent on $c_\mathrm{Br}$), and $\bm{f}(\bm{r})$ is the volume-force density. In cubic crystals it can be expressed using the gradient of $c_\mathrm{Br}$:
\beq{8}
f_j(\bm{r})=-\frac{\partial c_\mathrm{Br}(\bm{r})}{\partial x_j}(C_{11}+2C_{12})\eta;
\eeq
here we used the 2-index notation of the elastic constants $C_{11}\equiv C_{1111},\ C_{12}\equiv C_{1122}$.

In an infinite crystal, equations (\ref{e7},\ref{e8}) can be solved by Fourier method putting
\beq{9}
\bm{u}^{\mathrm{(FT)}}(\bm{k})=\int \D^3 \bm{r} \, \bm{u}(\bm{r}) \E^{-\I \bm{k}.\bm{r}},
\eeq
which converts the system of partial differential equations (\ref{e1}) to a set of linear algebraic equations that can be solved directly:
\beq{10}
\bm{u}^{\mathrm{(FT)}}(\bm{k})=\hat{\mathbf{G}}(\bm{k})\bm{f}^{\mathrm{(FT)}}(\bm{k}),
\eeq
where $\hat{\mathbf{G}}(\bm{k})$ is the elastic Green function defined as
\beq{11}
\left[\hat{\mathbf{G}}(\bm{k})^{-1}\right]_{jm}=C_{jkmn}k_k k_n.
\eeq
In all formulas the superscript $^\mathrm{(FT)}$ denotes Fourier transformation (FT), the Fourier transformation back to real space was carried out by a fast-Fourier numerical routine (FFT).

In the cubic symmetry, the elastic energy per one Br/I site is
\begin{equation}
	\label{e12}
	\begin{split}
		E_\mathrm{deformation}=\frac{1}{2\rho}\left\{C_{11} \left[ (\epsilon_{11}-\epsilon_0)^2 + (\epsilon_{22}-\epsilon_0)^2 +(\epsilon_{33}-\epsilon_0)^2\right] + \right.\\
		\left. + 2C_{12} \left[(\epsilon_{11}-\epsilon_0)(\epsilon_{22}-\epsilon_0)+ (\epsilon_{11}-\epsilon_0)(\epsilon_{33}-\epsilon_0)+ \right.\right.\\ \left.\left.+(\epsilon_{22}-\epsilon_0)(\epsilon_{33}-\epsilon_0)\right] + 4C_{44}\left( \epsilon_{12}^2+\epsilon_{23}^2+\epsilon_{13}^2\right) \right\},
	\end{split}
\end{equation}
where
\[
\epsilon_{jk}(\bm{r})=\frac{1}{2}\left[ \frac{\partial u_j(\bm{r})}{\partial x_k} + \frac{\partial u_k(\bm{r})}{\partial x_j}\right]
\]
is the strain tensor defined with respect to the lattice with $c_\mathrm{Br}=c_0$ and $\rho$ is the density of the Br/I sites in the MHP lattice ($3 \times$ number of MHP molecules per volume unit). The functional derivation of $E_\mathrm{deformation}$ yields
\beq{13}
\frac{\delta E_\mathrm{deformation}}{\delta \varphi}=\frac{\eta}{\rho}(C_{11}+2C_{12})(3\epsilon_0-\epsilon_{11}-\epsilon_{22}-\epsilon_{33}),
\eeq
this formula is used in the Cahn-Hilliard equation.

\section{Free carriers}

The first step in the simulation of the effect of free carriers consists in the calculation of the densities $n(\bm{r},t),\ p(\bm{r},t)$ of electrons in the conduction band and holes in the valence band for a given local Br content $c_\mathrm{Br}(\bm{r})$. The space fluctuations of the Br content result in fluctuations of the bandgap $E_g$, with increasing Br content and therefore with increasing $\varphi$, $E_g$ increases. In the simulation we assume that the generation rate $G$ (in m$^{-3}$s$^{-1}$, i.e., the number of electron-hole pairs generated in a unit volume during a unit time interval) is homogeneous in the whole simulation volume. The recombination rate $R$ in m$^{-3}$s$^{-1}$ includes both the bi-molecular and Schottky-Read-Hall (SRH) mono-molecular recombination processes, which gives \cite{FeldmannAOM2021}
\beq{14}
R=\frac{np}{\tau_n p +\tau_p n}+k_2np,
\eeq
where $\tau_{n,p}$ are the electron and hole lifetimes for the SRH recombination, and $k_2$ is the bi-molecular recombination constant. The continuity equations for electrons and holes are \cite{KirchartzAEM2020,FeldmannAOM2021}
\beq{15}
\begin{aligned}
	\frac{\D n}{\D t} & = D_n \Delta n +G -R,\\
	\frac{\D p}{\D t} & = D_p \Delta n -\mu_p \nabla\cdot(p\bm{E}) +G -R.
\end{aligned}
\eeq
Here we denoted $D_{n,p}$ the diffusion constants of electrons and holes, $\mu_{p}$ is the hole mobility. The electric field caused by the fluctuations of the bandgap $E_g(\bm{r})$ is
\[
\bm{E}=-\nabla E_g(\bm{r})=-\frac{\D E_g}{\D c_\mathrm{Br}}\,\nabla c_\mathrm{Br}(\bm{r}) ,
\]
and it acts as a driving force for the drift holes. Since the bandgap variation mostly affects the top of the valence band and the bottom of the conduction band is nearly constant, the drift of electrons in the conduction band can be neglected. In the continuity equations we do not consider the electric field $\Delta V = -e[p(\bm{r})-n(\bm{r})]/\varepsilon$ due to inhomogeneity of the carrier density, since the static permittivity $\varepsilon$ is large ($\varepsilon \approx 10^2 \div 10^3$) so that the electrostatic potential caused by the ion demixing is much stronger. 

In the following we assume that the motion of carriers is much faster than the ion diffusion, therefore the equilibrium carrier distribution is set almost instantaneously and it immediately follows the relatively slow changes of the ion distribution (adiabatic approximation). Therefore, in solving the continuity equations (\ref{e15}) we look for the stationary solution $\D n/\D t =\D p/\D t=0$ for a given ion distribution $c_\mathrm{Br}(\bm{r})$. In the next chapter we show that using reasonable estimates of the numerical values of constants occurring in these equations the equilibrium is reached in a few ns. 

For the derivation of the contribution $E_\mathrm{carriers}$ to the free energy we consider two microscopic models; model A calculates the total energy of free carriers in the simulation domain using the local bandgap $E_g(\varphi(\bm{r}))$, in model B we consider a local distortion of the lattice due to free carriers, which facilitates the ion demixing.

In model A, the energy of charge carriers $E_\mathrm{carriers}$ per one Br/I site can be calculated using the simple formula
\beq{16}
E_\mathrm{carriers}^{(\mathrm{A})}=\frac{1}{\rho V}\int_V \D^3\bm{r}\, p(\bm{r}) E_g(\varphi(\bm{r})).
\eeq
After some algebra we obtain the following formula for the functional derivative
\beq{18}
\frac{\delta E_\mathrm{carriers}^{\mathrm{(A)}}}{\delta \varphi}=\frac{1}{\rho V}\left(p \frac{\D E_g}{\D \varphi} +\frac{\partial p}{\partial \varphi} E_g \right).
\eeq
The electron density $n(\bm{r})$ exhibits much smaller local fluctuations than the density of holes, so that the influence of electrons can be neglected in the formula for $E_\mathrm{carriers}^{(\mathrm{A})}$. 

Model B assumes that the holes cause additional deformation of the lattice with the displacement
\beq{19}
\bm{u}_\mathrm{carriers}(\bm{r})=-\frac{\alpha}{4\pi \varepsilon_0}\int \D^3 \bm{r}\,  p(\bm{r'})\, \frac{\bm{r}-\bm{r'}}{\left|\bm{r}-\bm{r'}\right|^3},
\eeq
where $\varepsilon_0$ is the vacuum permittivity and $\alpha$ is the polarizability of the MHP lattice. In this model, the contribution of the holes to the total free energy is expressed using the deformation potential $U$ \cite{BischakNANO2017}:
\beq{20}
E_\mathrm{carriers}^{(\mathrm{B})}=\frac{U}{\rho V} \int_V \D^3\bm{r}\, p(\bm{r}) \, \left[\nabla.\bm{u}(\bm{r})\right].
\eeq
For the functional derivative we obtain after some calculation 
\beq{21}
\frac{\delta E_\mathrm{carriers}^{\mathrm{(B)}}}{\delta\varphi}=\frac{U \eta}{\rho}\frac{C_{11}+2C_{12}}{C_{11}}\left(p+\varphi\frac{\partial p}{\partial \varphi}\right)
\eeq

\section{Numerical simulations}

The CH equation (\ref{e1}) is solved using FT to reciprocal space \cite{LeeCMS2014}:
\beq{22}
\frac{\partial\varphi^\mathrm{(FT)}(\bm{k},t)}{\partial t}=-k^2 M \left[\Psi(\bm{k},t) + \kappa k^2 \varphi^\mathrm{(FT)}(\bm{k},t)\right], \Psi(\bm{k},t)\equiv\left(\frac{\delta F(\varphi(\bm{r},t))}{\delta \varphi(\bm{r},t)}\right)^\mathrm{(FT)}.
\eeq
The consequence of the Fourier transformation are the periodic boundary conditions. 

In the numerical implementation we use the stabilizing scheme \cite{LeeCMS2014}
\beq{23}
\begin{split}
	\frac{\varphi^\mathrm{(FT)}(\bm{k},t_{n+1})-\varphi^\mathrm{(FT)}(\bm{k},t_{n})}{\Delta t} \approx M\left\{-k^2\Psi(\bm{k},t_n) - \kappa k^4  \varphi^\mathrm{(FT)}(\bm{k},t_{n+1}) - \right. \\
	\left. - 2k^2 \left[\varphi^\mathrm{(FT)}(\bm{k},t_{n+1}) - \varphi^\mathrm{(FT)}(\bm{k},t_{n})\right]\right\},
\end{split}
\eeq
where $\Delta t= t_{n+1}-t_n$ is the time step. In the simulation of $\Psi(\bm{k},t)$ the FT of $\delta E_\mathrm{deformation}/\delta \varphi$ can be calculated directly, since it linearly depends on $\varphi$, while FTs of $\delta F_\mathrm{mix}/\delta \varphi$  and $\delta E_\mathrm{carriers}/\delta \varphi$ are performed numerically using a FFT routine.

The numerical solution of the CH equation is carried out in a three-dimensional orthogonal grid with typically $100 \times 100 \times 100$ grid points, with the step $\Delta{}x=10$\:{}\AA, assuming periodic boundary conditions. We define a random initial distribution of the order parameter $\varphi(\bm{r},t=0)$ with $\langle \varphi \rangle=0$, given rms deviation $\sigma_0=\sqrt{\langle \varphi^2\rangle} \ll 1$ and given correlation length $w$, the mean Br concentration was $c_0=0.4$. We calculated the evolution of the order parameter $\varphi$ in the time range from 0 to 1000\:{}ns; in the case of carrier generation we have chosen ten times shorter time steps to be able to follow a rapid onset of segregation. We have tested the accuracy and stability of the integration scheme (\ref{e23}) for various time steps $\Delta t$ and achieved good results for $\Delta t=0.1$\:{}ns and 0.01\:{}ns without and with carrier generation, respectively.

The solution constant $\Omega$ was calculated ab-initio using a DFT approach and it came out that it slightly depends on the local Br concentration $c_\mathrm{Br}$ \cite{BrivioJPCL2016}:
\[
\Omega(c_\mathrm{Br}) \approx \Omega_0 +\Omega_1 c_\mathrm{Br},
\]
where $\Omega_0\approx 0.06$\:{}eV/site, and $\Omega_1 \approx -0.02$\:{}eV/site. The interface constant $\kappa$ is not known. Its very rough estimate is based on the value $\eta\approx -0.065$ and on the elastic constants $C_{jk}$ which are of the order of 0.1\:{}eV/\AA$^3$. This gives a rough estimate of $\kappa$ between 0.05 and 0.5\:{}eV\AA$^2$/site. Fortunately, its value has almost no influence to the Br/I segregation kinetics and we set $\kappa=0.1$\:{}eV\AA$^2$/site as a typical value. The ion mobility $M$ is connected with the ion diffusion coefficient $D$ via the Einstein-Smoluchowski relation
\beq{24}
M=\frac{D}{k_BT},
\eeq
from which $M$ can be roughly estimated. The $D$ values in the literature show a large spread, typically $D \approx 10^{-8}$\:{}cm$^2$s$^{-1}\equiv{}10^8$\:{}\AA$^2$s$^{-1}$; from this value we can very roughly estimate $M\approx4\times10^9$\:{}\AA$^2$s$^{-1}$eV$^{-1}$ at room temperature. 

The dependence of the bandgap on the Br concentration can be approximated by the polynomial relation \cite{ChenNatComm2021}
\beq{25}
E_g(c_\mathrm{Br})=E_1(1-c_\mathrm{Br})+E_2 c_\mathrm{Br} +E_3 c_\mathrm{Br} (1-c_\mathrm{Br}),
\eeq
where $E_1=1/52$\:{}eV, $E_2=2.25$\:{}eV, and $E_3=-0.15$\:{}eV for FACsPbIBr perovskite. For the lifetimes $\tau_{n,p}$ and the bi-molecular recombination constant $k_2$ we used the literature values $\tau_n = 511$\:{}ns, $\tau_p=871$\:{}ns, and $k_2=8.1\times{}10^{-11}$\:{}cm$^3$/s \cite{RichterNatComm2016,StaubPRA2016}. The electron and hole mobilities are approximately $\mu_p \approx \mu_n \approx 30$\:{}cm$^2$V$^{-1}$s$^{-1}$ \cite{BiewaldAMI2019}, the carrier diffusion constant is $D_n\approx D_p = \mu_p k_B T/e \approx 0.78$\:{}cm$^2$s$^{-1}$.

For the deformation potential we used the literature value of $U=3.9\times 10^{-4}$\:{}eV/site \cite{BischakNANO2017}, the polarizability was calculated from the relative static permittivity $\varepsilon$ using the Claussius-Mossotti equation
\[
\frac{\varepsilon-1}{\varepsilon+2}=\frac{\alpha\rho}{3\varepsilon_0}.
\]
Since the static permittivity of a MHP crystal is rather high ($\varepsilon \approx 10^2 \div 10^3$ \cite{YangAngewChem2015}), the polarizability can be estimated as $\alpha \approx 3\varepsilon_0/\rho$.

\section{Simulation results}

\begin{figure}
	\includegraphics[width=15cm]{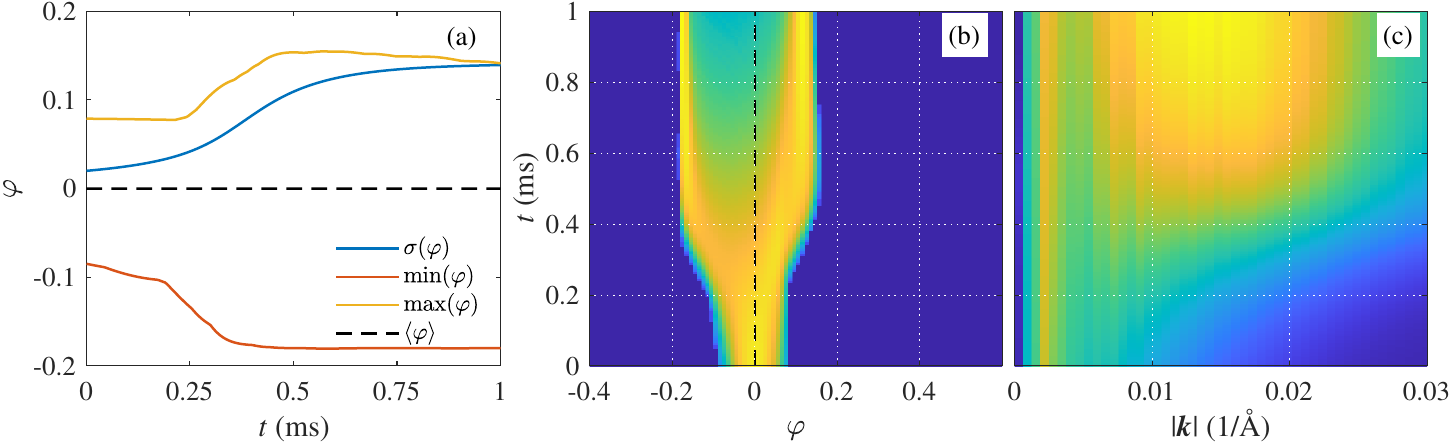}
	\caption{Examples of the simulation results calculated for $\sigma_0=0.02$, $w=20$\:{}nm, and $T=300$\:{}K, no elastic deformation and no free carrier generation ($\eta=0,\ G=0$): (a) The rms deviation $\sigma$ of the order parameter $\varphi$, its maximum and minimum values in the simulation domain, as well as its mean value $\langle\varphi \rangle$ as functions of time $t$; (b) histograms of the values of $\varphi(\bm{r},t)$; (c) azimuthally averaged values of $|\varphi^\mathrm{(FT)}(\bm{k},t)|$.}
	\label{f1}
\end{figure}

In the first step we calculate the ordering function $\varphi(\bm{r},t)$ without elastic deformation and without free carriers, i.e., we put $\eta=0,\ G=0$. Figure \ref{f1} shows the results for $\sigma_0=0.02$, $w=20$\:{}nm, and $T=300$\:{}K. In panel (a) we plotted the time dependences of the rms deviation $\sigma(t)=\sqrt{\langle \varphi(\bm{r},t)^2\rangle}$, as well as the minimum and maximum values of $\varphi(\bm{r},t)$, and its mean value $\langle \varphi(\bm{r},t)\rangle$ averaged over the simulation domain (dashed black line).
Since the solution of the CH equation obeys the conservation law of the total number of atoms, $\langle \varphi\rangle$ must be zero, and possible deviations from zero indicate numerical errors in the CH solution. Panel (b) shows the values of the histogram of $\varphi$, in (c) we plot the Fourier transformation $\varphi^\mathrm{(FT)}(\bm{k},t)=\int\D^3\bm{r} \varphi(\bm{r},t)\exp(\I \bm{k}.\bm{r})$ averaged over all directions of $\bm{k}$. The simulation results demonstrate a strong demixing which results in domains with Br concentrations of approx. 21\% and 52\%, which occur after 0.5\:{}ms. From panel (c) it follows that the domains are quite small with the characteristic size of about 120\:{}nm.

\begin{figure}
	\includegraphics[width=10cm]{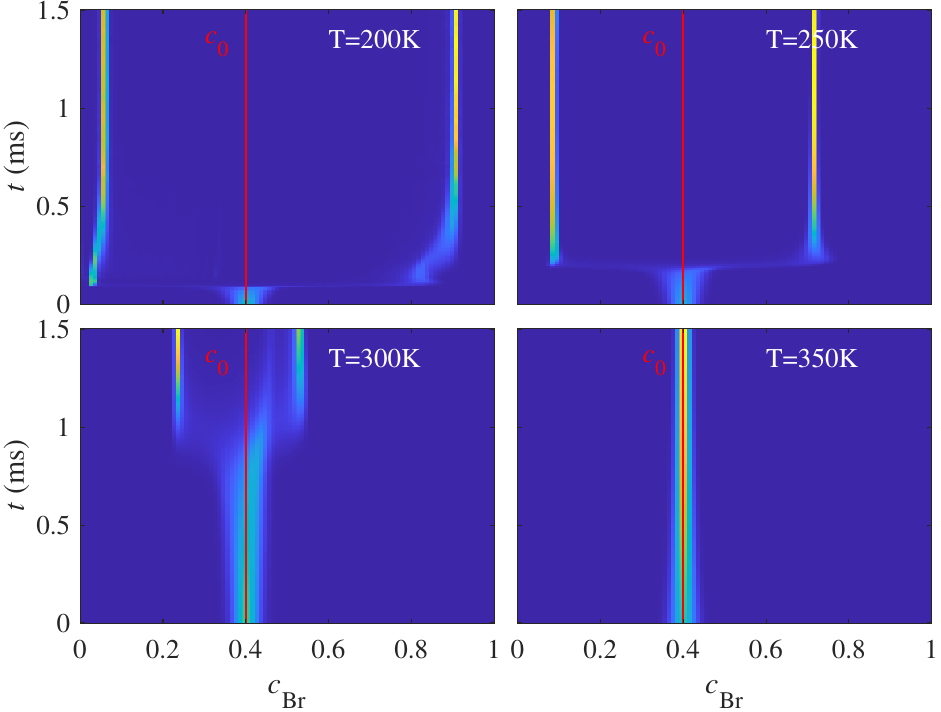}
	\caption{Time evolutions of the histograms of $c_\mathrm{Br}$ calculated for various temperatures $T$ with no elastic deformation and no free carriers. The vertical red lines represent the average Br concentration $c_0$.}
	\label{f9}
\end{figure}

\begin{figure}
	\includegraphics[width=10cm]{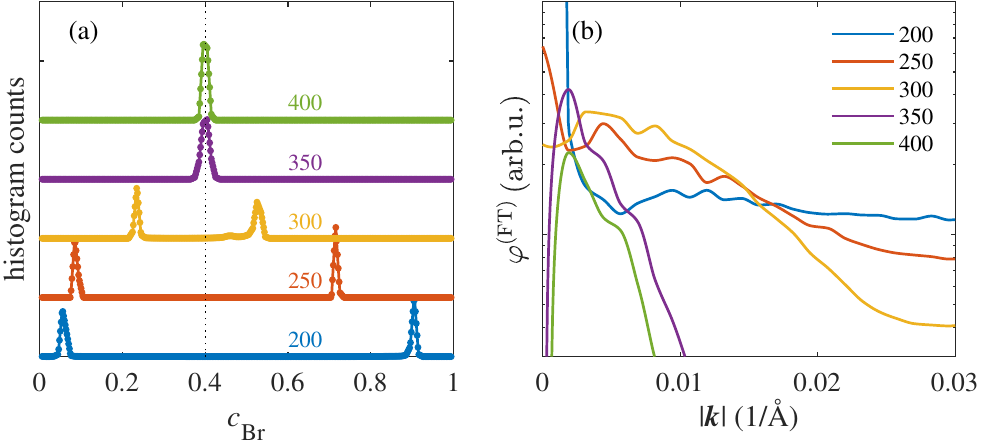}
	\caption{(a) Histograms of the values of $\varphi(\bm{r})$ calculated for various temperatures (the parameters of the curves in K), no elastic deformation and no free carrier generation. In panel (b) we show the azimuthally averaged values of $|\varphi^\mathrm{(FT)}(\bm{k},t)|$ calculated for various temperatures. In both panels the data for $t=1$\:{}ms are displayed.}
	\label{f2}
\end{figure}

The demixing effect is strongly temperature dependent; in Figure \ref{f9} we display the time evolution of the histograms of $c_\mathrm{Br}$ calculated for various temperatures $T$ without elastic deformation and without the free carriers. It is obvious that with increasing temperature the onset of the demixing process is shifted to later times and the final Br concentrations in the Br-rich and I-rich domains get closer to the mean concentration $c_0$. No demixing is obtained for $T>300$\:{}K. Fig. \ref{f2}(a) shows the histograms calculated for various temperatures and time $t=1$\:{}ms, in panel (b) we show the azimuthally averaged Fourer transformations of $\varphi(\bm{r})$ from the figure it follows that with increasing temperature the characteristic size of the fluctuations increases from 50\:{}nm to approx 300\:{}nm.

\begin{figure}[h]
	\includegraphics[width=10cm]{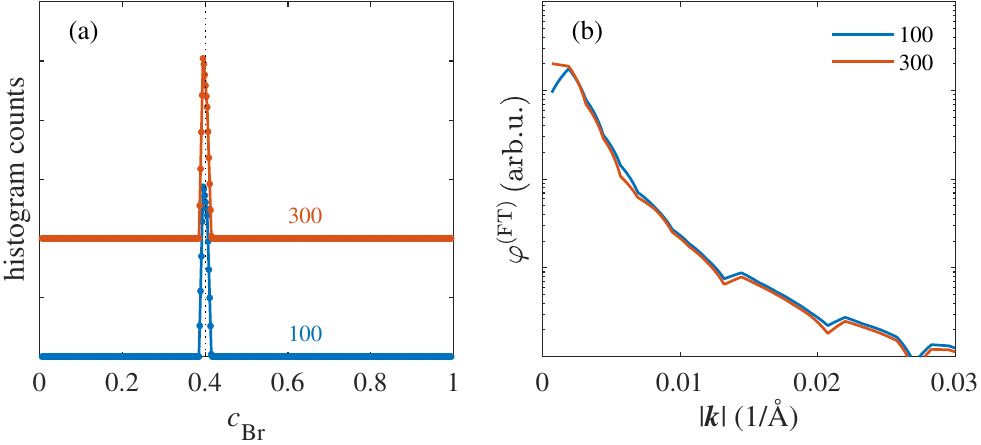}
	\caption{The same situation as in Fig. \ref{f2}, the simulation was performed including the energy of elastic deformation caused by the demixing.}
	\label{f3}
\end{figure}

Now we include the elastic deformation caused by the ion demixing. Figure \ref{f3} demonstrates that even at a very low temperature of $T=100$\:{}K the histogram of the $\varphi(\bm{r})$ values shows only a narrow peak around $c_0$. Therefore, the elastic deformation caused by Br/I fluctuations suppresses effectively the demixing, since the elastic energy of the deformation is too large.

\begin{figure}[h]
	\includegraphics[width=7cm]{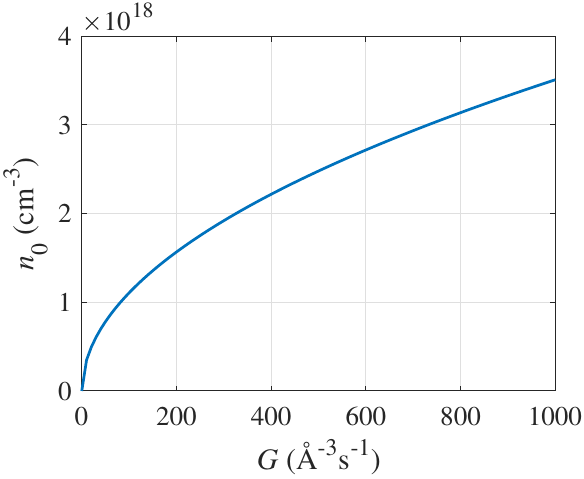}
	\caption{Dependence of the equilibrium carrier density $n_0$ on the generation rate $G$.}
	\label{f10}
\end{figure}

The first step in the simulation of the influence of free carriers is the solution of the continuity equations (\ref{e15}). The starting condition of the simulation run was assumed in the form
\[
n(\bm{r},0)=p(\bm{r},0)=n_0,
\]
where $n_0$ is the equilibrium homogeneous density of free carriers at the given generation rate, following from the condition $G=R$:
\[
n_0=\frac{1}{2k_2}\left(-\frac{1}{\tau}+\sqrt{\frac{1}{\tau^2}+4k_2G}\right),\, \tau=\tau_n+\tau_p.
\]
Figure \ref{f10} shows the dependence of this equilibrium density on the generation rate $G$, the graph demonstrates that reasonable carrier densities of several $10^{18}$\:{}cm$^{-3}$ can be reached for the generation rate in hundreds of \AA$^{-3}$s$^{-1}$. 

\begin{figure}[h]
	\includegraphics[width=15cm]{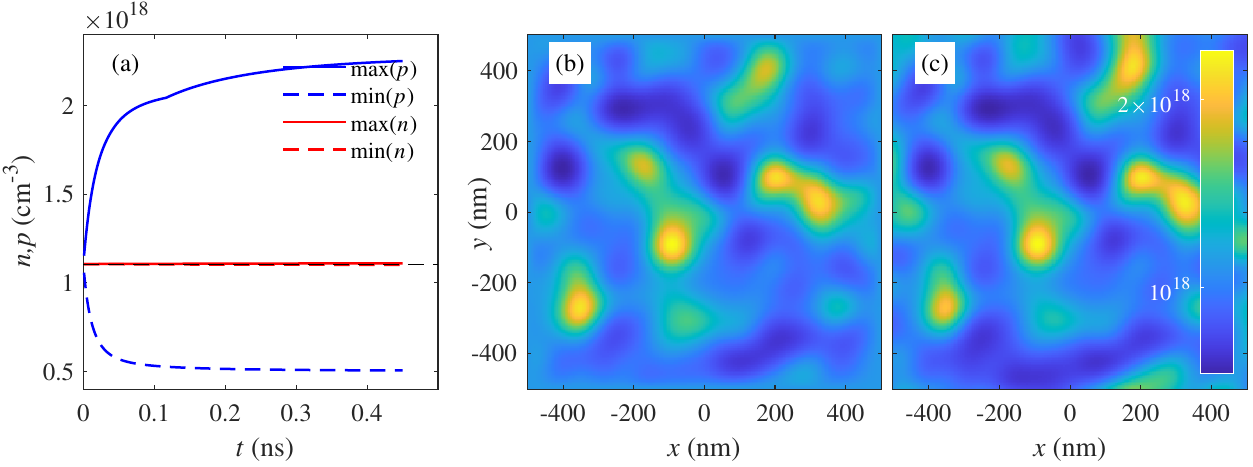}
	\caption{(a) The time dependence of the maximum and minimum values of the hole and electron densities $p(\bm{r},t),\ p(\bm{r},t)$ calculated for the generation rate $G=100$\:{}\AA$^{-3}$s$^{-1}$, the rms deviation of the Br concentration was $\sigma_0=0.01$ and the correlation length $w=100$\:{}nm. In panels (b) and (c) the final density $p(\bm{r})$ and the equilibrium density $p_\mathrm{eq}(\bm{r})$ are plotted, respectively; the units in the colorbar are cm$^{-3}$. }
	\label{f4}
\end{figure}

In Figure \ref{f4} we plot the results of the numerical solution of the drift-diffusion equations (\ref{e15}), in the simulations we set the generation rate $G=100$\:\AA$^{-3}$s$^{-1}$. In panel (a) we show the time dependence of the maximum and minimum values of the hole and electron concentrations $p(\bm{r},t),\ n(\bm{r},t)$, the electron density is almost homogeneous and its fluctuation from the value of $n_0$ are negligible. From (a) it is obvious that the hole distribution stabilizes at approx. 0.5\:{}ns. Panel (b) displays the final distribution of the hole density. For comparison, panel (c) displays the equilibrium hole concentration calculated in the quasiclassical approximation
\beq{26}
p_\mathrm{eq}(\bm{r})=n_0 \frac{V}{Z}\exp\left[-\frac{E_g(\bm{r})}{k_BT}\right], Z=\int_V \D^3\bm{r}\exp\left[-\frac{E_g(\bm{r})}{k_BT}\right],
\eeq
where $V$ is the simulation volume. The hole distribution after approx 0.5\:{}ns is very similar to $p_\mathrm{eq}(\bm{r})$ so that we use $p(\bm{r})=p_\mathrm{eq}(\bm{r})$ calculated from the current $E_g(\bm{r})$ distribution in each time step of the demixing simulation run.

\begin{figure}[h]
	\includegraphics[width=10cm]{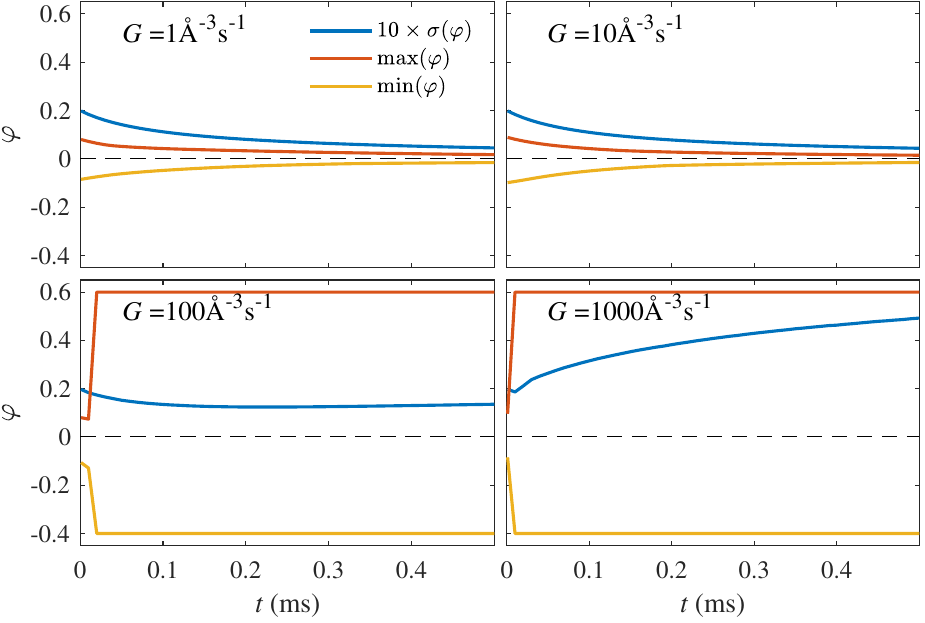}
	\caption{The rms deviation $\sigma$, and the maximum and minimum values of the order parameter $\varphi$ calculated for various generation rates $G$ and $\sigma_0=0.02,\, w=50$\:{}nm, model A.}
	\label{f5}
\end{figure}

\begin{figure}[h]
	\includegraphics[width=10cm]{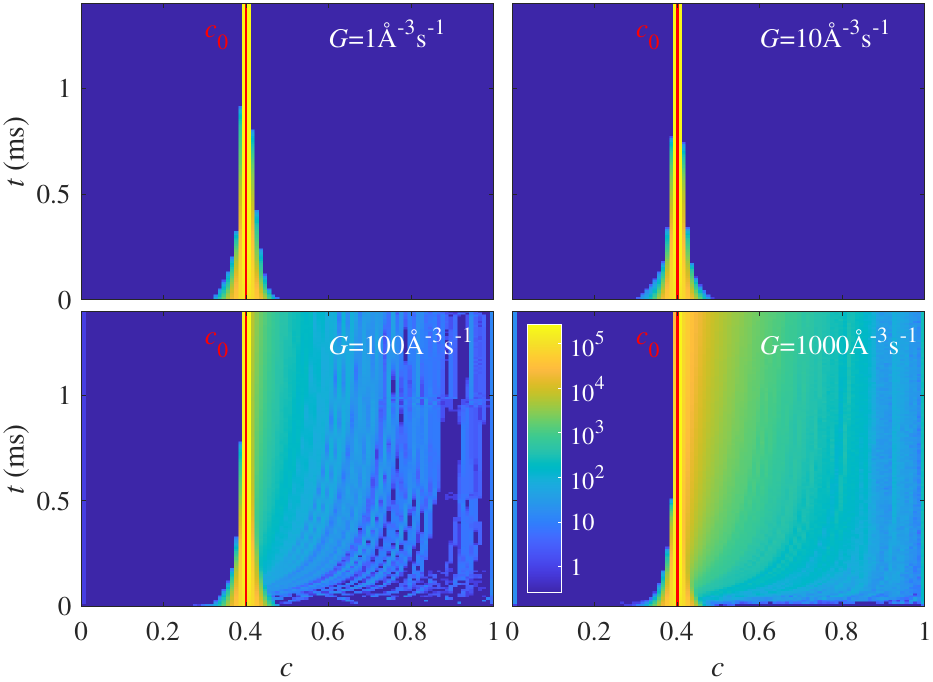}
	\caption{Time evolution of the histograms of $c_\mathrm{Br}$ calculated for various generation rates, and $\sigma_0=0.02,\, w=50$\:{}nm, model A.}
	\label{f6}
\end{figure}

We performed a series of simulations assuming the model A mentioned above, here we present typical examples. Figures \ref{f5} and \ref{f6} show the time dependence of rms deviations $\sigma$, maximum and minimum values of $\varphi$, as well as the time evolution of the histogram of $c_\mathrm{Br}(\bm{r},t)$, calculated for various generation rates, assuming the starting deviation $\sigma_0=0.02$ and the correlation length $w=50$\:{}nm. The figures demonstrates that the demixing is observed only for rather big generation rates of the order of $G=100$\:{}\AA$^{-3}$s$^{-1}$ and larger; these rates correspond to the homogeneous carrier density $n_0 \approx 10^{18}$\:{}cm$^{-3}$. The ion distribution resulting from this demixing process substantially differs from the demixing without free carriers described above (figures \ref{f1} and \ref{f2}). The demixing driven by free carriers results in pure Br- and I-containing domains, corresponding to $c_\mathrm{Br}=1$ and 0, respectively (i.e., $\varphi=0.6$ and $\varphi=-0.4$), while the demixing without free carriers (driven solely by $F_\mathrm{mix}$) creates Br-rich and I-rich domains with temperature-dependent chemical composition (see Figures \ref{f9} and \ref{f2}).

\begin{figure}[h]
	\includegraphics[width=15cm]{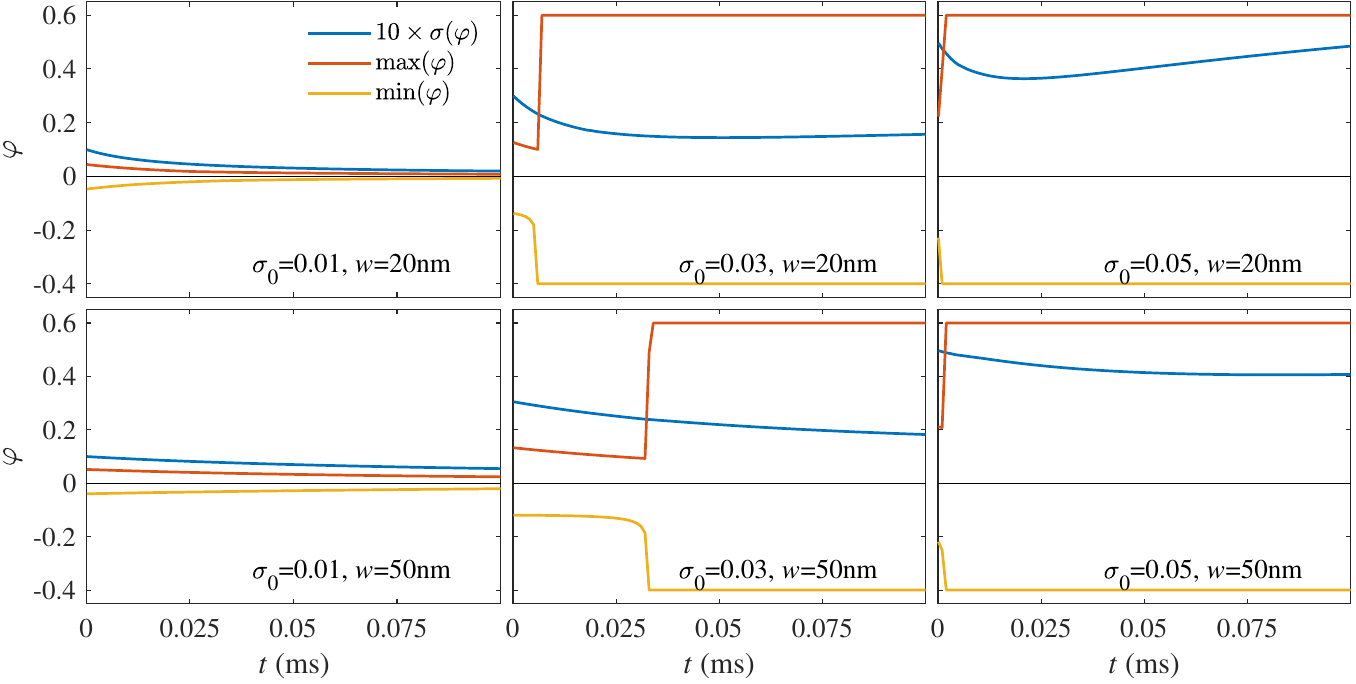}
	\caption{The rms deviation $\sigma$, and the maximum and minimum values of the order parameter $\varphi$ calculated for the same generation rate $G=100$\:{}\AA$^{-3}$s$^{-1}$ and various $\sigma_0$ and $w$, model A.}
	\label{f7}
\end{figure}

In contrast to the demixing without free carriers, the resulting structure also strongly depends on the starting conditions, i.e., on the values $\sigma_0$ and $w$ (Figure \ref{f7}). The figure shows that the onset of the demixing appears earlier for larger $\sigma_0$ and smaller $w$. However, the exact onset time is random and it depends on the particular starting distribution $c_\mathrm{Br}(\bm{r},t=0)$. This is more pronounced in the case of larger correlation lengths, for which the ergodicity of the simulation is not fully ensured.

\begin{figure}[h]
	\includegraphics[width=15cm]{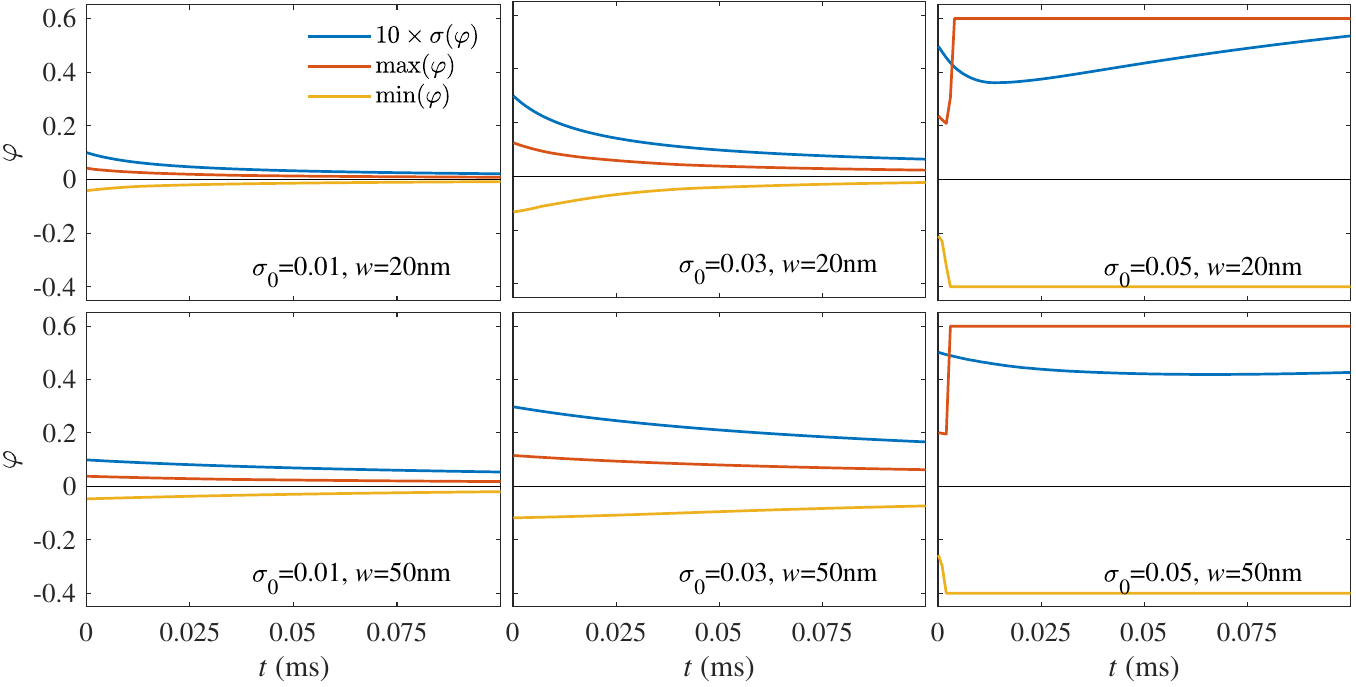}
	\caption{The same situation as in Figure \ref{f7}, model B.}
	\label{f8}
\end{figure}

The simulation results for model B are qualitatively very similar, in this model however the demixing starts only for larger starting rms deviation $\sigma_0$. This shows Figure \ref{f8} showing that for $G=100$\:{}\AA$^{-3}$s$^{-1}$ the demixing is observed only for $\sigma_0=0.05$ and larger.

%\section{Discussion}

\section{Summary}

The results of the numerical simulations can be summarized in the following items:
\begin{enumerate}
	\item
	Without elastic deformation and without free carriers the demixing is driven only by the chemical part $F_\mathrm{mix}$ of the free energy. The chemical compositions of the Br-rich and I-rich domains depend on temperature, with increasing temperature the difference of the compositions decreases and no demixing is observed above 300\:{}K.
	\item
	Elastic deformation caused by the dependence of the lattice parameter on the Br concentration suppresses the demixing at all temperatures.
	\item
	Ion demixing caused by free carriers (bandgap fulctuations -- model A , or polaron model -- model B) creates \emph{pure} Br- and I-containing domains ($c_\mathrm{Br}=0$ or 1).
	\item
	Free-carriers-driven demixing depends on starting conditions, in particular the demixing appears only for starting values of the rms deviation $\sigma$ of the Br concentration larger than approx. 0.03 and 0.05 for models A and B, respectively.
\end{enumerate}

\section*{Acknowledgement}
The work has been supported by the Czech Science Foundation (project 23-06543S).

\bibliography{references}

\end{document}